\documentclass[twocolumn]{aastex63}

\newcommand\teff{\mbox{$T_\mathrm{eff}$}}
\newcommand\logg{\mbox{$\log{g}$}}
\usepackage{color}

\begin{document}

\title{Redder than Red: Discovery of an Exceptionally Red L/T Transition Dwarf}

\correspondingauthor{Adam C. Schneider}
\email{aschneid10@gmail.com}

\author[0000-0002-6294-5937]{Adam C. Schneider}
\affil{United States Naval Observatory, Flagstaff Station, 10391 West Naval Observatory Rd., Flagstaff, AZ 86005, USA}

\author[0000-0002-6523-9536]{Adam J. Burgasser}
\affil{Center for Astrophysics and Space Science, University of California San Diego, La Jolla, CA 92093, USA}

\author[0000-0002-3858-1205]{Justice Bruursema}
\affil{United States Naval Observatory, Flagstaff Station, 10391 West Naval Observatory Rd., Flagstaff, AZ 86005, USA}

\author[0000-0002-4603-4834]{Jeffrey A. Munn}
\affil{United States Naval Observatory, Flagstaff Station, 10391 West Naval Observatory Rd., Flagstaff, AZ 86005, USA}

\author{Frederick J. Vrba}
\affil{United States Naval Observatory, Flagstaff Station, 10391 West Naval Observatory Rd., Flagstaff, AZ 86005, USA}

\author[0000-0001-7896-5791]{Dan Caselden}
\affil{Department of Astrophysics, American Museum of Natural History, Central Park West at 79th St., New York, NY 10024, USA}

\author[0000-0003-4905-1370]{Martin Kabatnik}
\affil{Backyard Worlds: Planet 9, USA}

\author[0000-0003-4083-9962]{Austin Rothermich}
\affil{Department of Astrophysics, American Museum of Natural History, Central Park West at 79th St., New York, NY 10024, USA}

\author[0000-0003-4864-5484]{Arttu Sainio}
\affil{Backyard Worlds: Planet 9, USA}

\author[0000-0003-2235-761X]{Thomas P. Bickle}
\affil {School of Physical Sciences, The Open University, Milton Keynes, MK7 6AA, UK}
\affil {Backyard Worlds: Planet 9, USA}

\author[0000-0002-2968-2418]{Scott E. Dahm}
\affil{United States Naval Observatory, Flagstaff Station, 10391 West Naval Observatory Rd., Flagstaff, AZ 86005, USA}

\author[0000-0002-1125-7384]{Aaron M. Meisner}
\affil{NSF's National Optical-Infrared Astronomy Research Laboratory, 950 N. Cherry Ave., Tucson, AZ 85719, USA}

\author[0000-0003-4269-260X]{J. Davy Kirkpatrick}
\affil{IPAC, Mail Code 100-22, Caltech, 1200 E. California Blvd., Pasadena, CA 91125, USA}

\author[0000-0002-2011-4924]{Genaro Su{\'a}rez}
\affil{Department of Astrophysics, American Museum of Natural History, Central Park West at 79th St., New York, NY 10024, USA}

\author[0000-0002-2592-9612]{Jonathan Gagn\'e}
\affil{Plan\'etarium Rio Tinto Alcan, Espace pour la Vie, 4801 ave. Pierre-de Coubertin, Montr\'eal, QC H1V~3V4, Canada}
\affil{Institute for Research on Exoplanets, Universit\'e de Montr\'eal, 2900 Boulevard \'Edouard-Montpetit Montr\'eal, QC H3T~1J4, Canada}

\author[0000-0001-6251-0573]{Jacqueline K. Faherty}
\affil{Department of Astrophysics, American Museum of Natural History, Central Park West at 79th St., New York, NY 10024, USA}

\author[0000-0003-0489-1528]{Johanna M. Vos}
\affil{Department of Astrophysics, American Museum of Natural History, Central Park West at 79th St., New York, NY 10024, USA}

\author[0000-0002-2387-5489]{Marc J. Kuchner}
\affil{Exoplanets and Stellar Astrophysics Laboratory, NASA Goddard Space Flight Center, 8800 Greenbelt Road, Greenbelt, MD 20771, USA}

\author[0000-0002-3858-1205]{Stephen J. Williams}
\affil{United States Naval Observatory, Flagstaff Station, 10391 West Naval Observatory Rd., Flagstaff, AZ 86005, USA}

\author[0000-0001-8170-7072]{Daniella Bardalez Gagliuffi}
\affil{Department of Physics \& Astronomy, Amherst College, 25 East Drive, Amherst, MA 01003, USA}

\author[0000-0003-2094-9128]{Christian Aganze}
\affil{Center for Astrophysics and Space Science, University of California San Diego, La Jolla, CA 92093, USA}

\author[0000-0002-5370-7494]{Chih-Chun Hsu}
\affil{Center for Astrophysics and Space Science, University of California San Diego, La Jolla, CA 92093, USA}

\author[0000-0002-9807-5435]{Christopher Theissen}
\affil{Center for Astrophysics and Space Science, University of California San Diego, La Jolla, CA 92093, USA}

\author[0000-0001-7780-3352]{Michael C. Cushing}
\affil{Ritter Astrophysical Research Center, Department of Physics and Astronomy, University of Toledo, 2801 W. Bancroft St., Toledo, OH 43606, USA}

\author[0000-0001-7519-1700]{Federico Marocco}
\affil{IPAC, Mail Code 100-22, Caltech, 1200 E. California Blvd., Pasadena, CA 91125, USA}

\author[0000-0003-2478-0120]{Sarah Casewell}
\affil{School of Physics and Astronomy, University of Leicester, University Road, Leicester, LE1 7RH, UK}

\author{The Backyard Worlds: Planet 9 Collaboration}

\begin{abstract}

We present the discovery of CWISE J050626.96$+$073842.4 (CWISE J0506$+$0738), an L/T transition dwarf with extremely red near-infrared colors discovered through the Backyard Worlds: Planet 9 citizen science project.  Photometry from UKIRT and CatWISE give a $(J-K)_{\rm MKO}$ color of 2.97$\pm$0.03 mag and a $J_{\rm MKO}-$W2 color of 4.93$\pm$0.02 mag, making CWISE J0506$+$0738 the reddest known free-floating L/T dwarf in both colors.  We confirm the extremely red nature of CWISE J0506$+$0738 using Keck/NIRES near-infrared spectroscopy and establish that it is a low-gravity late-type L/T transition dwarf.  The spectrum of CWISE J0506$+$0738 shows possible signatures of CH$_4$ absorption in its atmosphere, suggesting a colder effective temperature than other known, young, red L dwarfs. We assign a preliminary spectral type for this source of L8$\gamma$--T0$\gamma$.  We tentatively find that CWISE J0506$+$0738 is variable at 3--5 $\mu$m based on multi-epoch WISE photometry. Proper motions derived from follow-up UKIRT observations combined with a radial velocity from our Keck/NIRES spectrum and a photometric distance estimate indicate a strong membership probability in the $\beta$ Pic moving group. A future parallax measurement will help to establish a more definitive moving group membership for this unusual object.  

\end{abstract}


\section{Introduction}
\label{sec:intro}

Young L dwarfs are often noted to be redder than their field-age counterparts \citep{kirkpatrick2008, cruz2009, faherty2013, faherty2016, liu2016}.  This is partly due to their lower surface gravities, which lead to lower pressures in their atmospheres, and hence reduced collision-induced absorption (CIA) by H$_2$ \citep{lin69}.  Further, lower surface gravities can lead to higher altitude clouds, leading to less efficient gravitational settling of condensate particles (e.g., \citealt{madhu2011, helling2014}).  Red near-infrared colors have been efficiently utilized to characterize and discover new young brown dwarfs and planetary-mass objects (e.g., \citealt{kellogg2015, schneider2017}).  There also exists a population of red L dwarfs that do not have obvious signs of youth (e.g., \citealt{looper2008, kirkpatrick2010, marocco2014}).  While the exact reasons for the red colors of these relatively high-gravity objects are not entirely clear, their spectra have been well-reproduced by the presence of micron or submicron-sized grains in their upper atmospheres \citep{marocco2014, hiranaka2016, charnay2018}.  This high-altitude dust suppresses emission at shorter wavelengths much more efficiently than longer wavelengths, leading to significantly reddened spectra compared to ``normal'' brown dwarfs.  There is evidence that the strength of silicate absorption features in the mid-infrared correlates with the near-infrared colors of L dwarfs \citep{burgasser2008,suarez2022}, indicating that variations in silicate cloud thickness also plays a role.  Further, viewing angle \citep{vos2017} and variability \citep{ashraf2022} have been shown to be related to the colors of substellar objects. There is also evidence that convective instabilities can produce similar effects as clouds in young red L dwarfs \citet{tremblin2017}.  In any case, young red L dwarfs and old reddened L dwarfs have proven to be compelling laboratories for the study of low temperature substellar atmospheres.  

The vast majority of the current population of directly-imaged planetary-mass companions are also young and have similar effective temperatures, masses, and radii as young L dwarfs, as well as observed properties, including unusually red near-infrared colors.  Examples include 2M1207b \citep{chauvin2004, chauvin2005, patience2010}, HD 206893B \citep{milli2017, delorme2017, krammerer2021, meshkat2021, ward2021}, VHS J125601.92$-$125723.9B \citep{gauza2015}, 2MASS J22362452$+$4751425b \citep{bowler2017},  BD$+$60 1417B \citep{faherty2021}, HR8799bcd \citep{marois2008}, and HD 203030B \citep{metchev2006}. Young, red L dwarfs in the field provide an opportunity to study the physical properties of giant exoplanet-like atmospheres without the technical challenge of blocking host star light.  

In this article, we present the discovery of CWISE J050626.96$+$073842.4 (CWISE J0506$+$0738), an exceptionally red brown dwarf discovered as part of the Backyard Worlds: Planet 9 (BYW) citizen science project \citep{kuchner2017}.  We detail its discovery in Section \ref{sec:discovery}, present Keck/NIRES spectroscopic follow-up observations in Section \ref{sec:obs}, analyze these data in Section \ref{sec:anal}, and discuss CWISE J0506$+$0738 in the context of other red brown dwarfs in Section \ref{sec:discussion}. 

\section{Discovery of CWISE 0506+0738}
\label{sec:discovery}

\begin{figure*}
\plotone{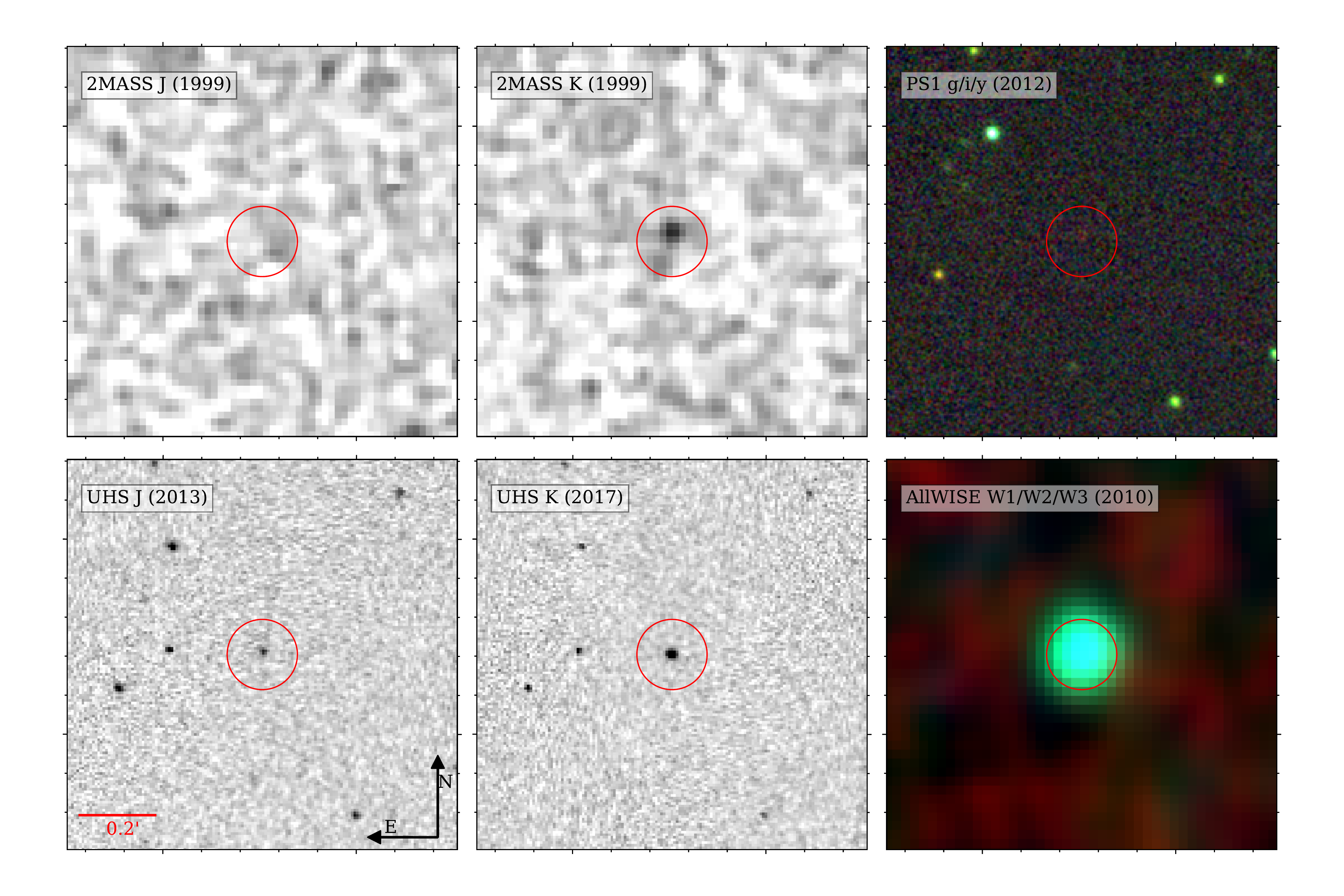}
\caption{Images of CWISE J0506$+$0738 from 2MASS (upper left and center), UHS (bottom left and center), Pan-STARRS (upper right, three-color image with $g/i/y$ bands), and WISE (lower right, three-color image with $W1/W2/W3$ bands).  The position of CWISE J0506$+$0738 as determined in the UHS $K$-band images is denoted by a red circle.  Note that CWISE J0506$+$0738 is undetected at 2MASS $J$ and in the Pan-STARRS 3-color image, but clearly detected in the 2MASS $K$-band, UHS, and WISE images.  The greenish hue of CWISE J0506$+$0738 in the WISE images shows that this object is significantly brighter at WISE channel W2 (4.6 $\mu$m) than WISE channel W1 (3.4 $\mu$m) or W3 (12 $\mu$m), typical of brown dwarfs with late-L or later spectral types.}  
\label{fig:finder}
\end{figure*}

CWISE J0506$+$0738 was submitted as an object of interest to the BYW project by citizen scientists Austin Rothermich, Arttu Sainio, Sam Goodman, Dan Caselden, and Martin Kabatnik because it had notable motion amongst epochs of WISE observations.   BYW uses unWISE images \citep{lang2014, meisner2018} covering the 2010--2016 time frame and is typically sensitive to objects with proper motions $\gtrsim$ 0\farcs05--0\farcs1 yr$^{-1}$.  As part of the initial investigation to evaluate whether or not CWISE J0506$+$0738 was a newly discovered substellar object, we gathered available photometry from the Two Micron All-Sky Survey (2MASS) reject catalog \citep{skrutskie2006, tmass2006}, the United Kingdom Infrared Telescope (UKIRT) Hemisphere Survey DR1 (UHS; \citealt{dye2018}), and the CatWISE 2020 main catalog \citep{marocco2021}, and determined a photometric spectral type of $\sim$L7.5 using the method described in \cite{schneider2016a}.   It was noted during the initial evaluation of this object that its $J-K$ color, using UHS $J$- and 2MASS $K$-band photometry, was exceptionally red ($J-K$ = 3.17$\pm$0.21 mag), more than half a magnitude redder than the reddest known free-floating L dwarf, PSO J318.5338$-$22.8603 ($J-K$ = 2.64$\pm$0.02 mag; \citealt{liu2013}).  An inspection of 2MASS, UHS, WISE, and Pan-STARRS DR2 \citep{magnier2020} images showed no sources of contamination, suggesting that the near-infrared colors accurately reflect the true spectral energy distribution of the source (Figure \ref{fig:finder}). 

The astrometry and photometry of CWISE J0506$+$0738 were further analyzed using measurements from the UHS DR2 catalog, which will provide $K$-band photometry for much of the northern hemisphere (Bruursema et al.~in prep.).   CWISE J0506$+$0738 was found to have a $K$-band magnitude of 15.513$\pm$0.022 mag, consistent with the previous 2MASS measurement but significantly more precise.  This measurement results in a UHS $(J-K)_{\rm MKO}$ color of 3.24$\pm$0.10 mag, slightly redder but consistent with UHS and 2MASS photometry.  We therefore considered this candidate a high-priority target for follow-up spectroscopic observations.   

\begin{deluxetable}{lcccc}
\label{tab:cwise0506}
\tablecaption{Properties of CWISE J050626.96$+$073842.4}
\tablehead{
\colhead{Parameter} & \colhead{Value} & \colhead{Ref.}}
\startdata
R.A. (\degr) (epoch=2022.7)\tablenotemark{a} & 76.6124377 & 1 \\
Dec. (\degr) (epoch=2022.7)\tablenotemark{a} & 7.6449299 & 1 \\
R.A. (\degr) (epoch=2017.8)\tablenotemark{a} & 76.6123885 & 2 \\
Dec. (\degr) (epoch=2017.8)\tablenotemark{a} & 7.6450716 & 2 \\
$\mu$$_{\alpha}$ (mas yr$^{-1}$) & 31.5$\pm$2.6 & 1\\
$\mu$$_{\delta}$ (mas yr$^{-1}$)  & -82.7$\pm$2.7 & 1\\
$d$\tablenotemark{b} (pc) & 32$^{+4}_{-3}$ & 1 \\
RV (km s$^{-1}$) & +16.3$^{+8.8}_{-7.7}$ & 1 \\ 
$J_{\rm MKO}$ (mag) & 18.487$\pm$0.017 & 1 \\
$K_{\rm MKO}$ (mag) & 15.513$\pm$0.022 & 2 \\
W1 (mag) & 14.320$\pm$0.015 & 3 \\
W2 (mag) & 13.552$\pm$0.013 & 3 \\
Sp.~Type &  L8$\gamma$--T0$\gamma$ & 1 \\
\enddata
\tablenotetext{a}{R.A. and Dec. values are given in the ICRS coordinate system.}
\tablenotetext{b}{Photometric distance estimate based on the UHS $K_{\rm MKO}$-band magnitude and the absolute magnitude-spectral type relation in \citealt{dupuy2012} (see Section \ref{sec:dist}).}
\tablerefs{ (1) This work; (2) UHS DR2 (\citealt{dye2018}, Bruursema et al.~in prep); (3) CatWISE 2020 \citep{marocco2021} }
\end{deluxetable}

\section{Observations}
\label{sec:obs}

\subsection{UKIRT/WFCAM}
\label{sec:ukirt}

In an effort to refine the astrometry and photometry of CWISE J0506$+$0738, we observed it with the $J_{\rm MKO}$ filter on the infrared Wide-Field Camera (WFCAM; \citealt{casali2007}) on UKIRT on 20 September 2022.  Observations were performed using a 3 $\times$ 3 microstepping pattern, with the resulting 9 images interleaved \citep{dye2006} to provide improved sampling over that of a single WFCAM exposure.  The microstepping sequence was repeated five times, resulting in 45 single exposures each lasting 20 seconds, for a total exposure time of 900 seconds.   We re-registered the world coordinate system (WCS) of each interleaved frame using the Gaia DR3 catalog \citep{gaia2022}. Images were then combined using the \texttt{imstack} routine from the \texttt{CASUTOOLS} package\footnote{http://casu.ast.cam.ac.uk/surveys-projects/software-release} \citep{irwin2004}.  The position and photometry of CWISE J0506$+$0738 were extracted using the \texttt{CASUTOOLS} \texttt{imcore} routine.  

Combining the position of this $J$-band observation with the UHS $K$-band observation, we calculated proper motion components of $\mu$$_{\alpha}$ = 31.5$\pm$2.6 mas yr$^{-1}$ and $\mu$$_{\delta}$ = -82.7$\pm$2.7 mas yr$^{-1}$.  CatWISE 2020 reports proper motions of $\mu$$_{\alpha}$ = 44.2$\pm$7.9 mas yr$^{-1}$ and $\mu$$_{\delta}$ = -97.5$\pm$8.4 mas yr$^{-1}$ (with offset corrections applied according to \citealt{marocco2021}).  The proper motion calculated from our UKIRT observations is significantly more precise than the proper motion measurements from CatWISE 2020, and we adopt the former for our analysis.

We measure a $J_{\rm MKO}-$band magnitude of 18.487$\pm$0.017 mag from these observations, which is $>$2$\sigma$ brighter than the value from the UKIRT Hemisphere Survey (18.76$\pm$0.10 mag).  To verify our measured photometry, we compared the photometry for other sources found to have similar magnitudes (18.4 $< J <$ 18.6 mag) in our images to UHS values.  We found a median $J$-band difference for the 52 objects in this sample to be -0.03 mag, with a median absolute deviation of 0.07 mag, showing that differences as large as that measured for this object (0.27 mag) are relatively rare.  The origin of the difference between these $J$-band measurements is unclear, though we note that variability may be a contributing factor, as young (and red) objects are often found to have larger amplitude variability than field-age objects with similar spectral types (e.g., \citealt{vos2022}).  While this new $J$-band measurement results in bluer $(J-K)_{\rm MKO}$ = 2.97$\pm$0.03 mag and $J_{\rm MKO}-$W2 = 4.94$\pm$0.02 mag colors, they both remain significantly redder than those of any previously identified free-floating brown dwarf.

All UKIRT photometry and astrometry for CWISE J0506$+$0738 are provided in Table \ref{tab:cwise0506}.

\subsection{Keck/NIRES}

CWISE J0506$+$0738 was observed with the Near-Infrared Echellette Spectrometer (NIRES; \citealt{wilson2004}) mounted on the Keck II telescope on UT 19 January 2022.  NIRES provides a resolution  $\lambda/\Delta\lambda$ $\approx$ 2700 over five cross-dispersed orders spanning a wavelength range of 0.9--2.45 $\mu$m.  CWISE J0506$+$0738 was observed in four 250 second exposures nodded in an ABBA pattern along the slit, which was aligned with the parallactic angle, for a total on-source integration time of 1000 seconds.  The spectrum was extracted using a modified version of the SpeXTool package \citep{vacca2003, cushing2004}, with the A0~V star HD 37887 ($V$ = 7.67) used for telluric correction.  The large $J-K$ color of CWISE J0506$+$0738 resulted in significant signal to noise (S/N) differences across the final reduced spectrum, with a S/N$\sim$25 at the $J$-band peak ($\sim$1.3~$\mu$m) and a S/N$\sim$200 at the $K$-band peak ($\sim$2.2~$\mu$m).  

\begin{figure*}
\plotone{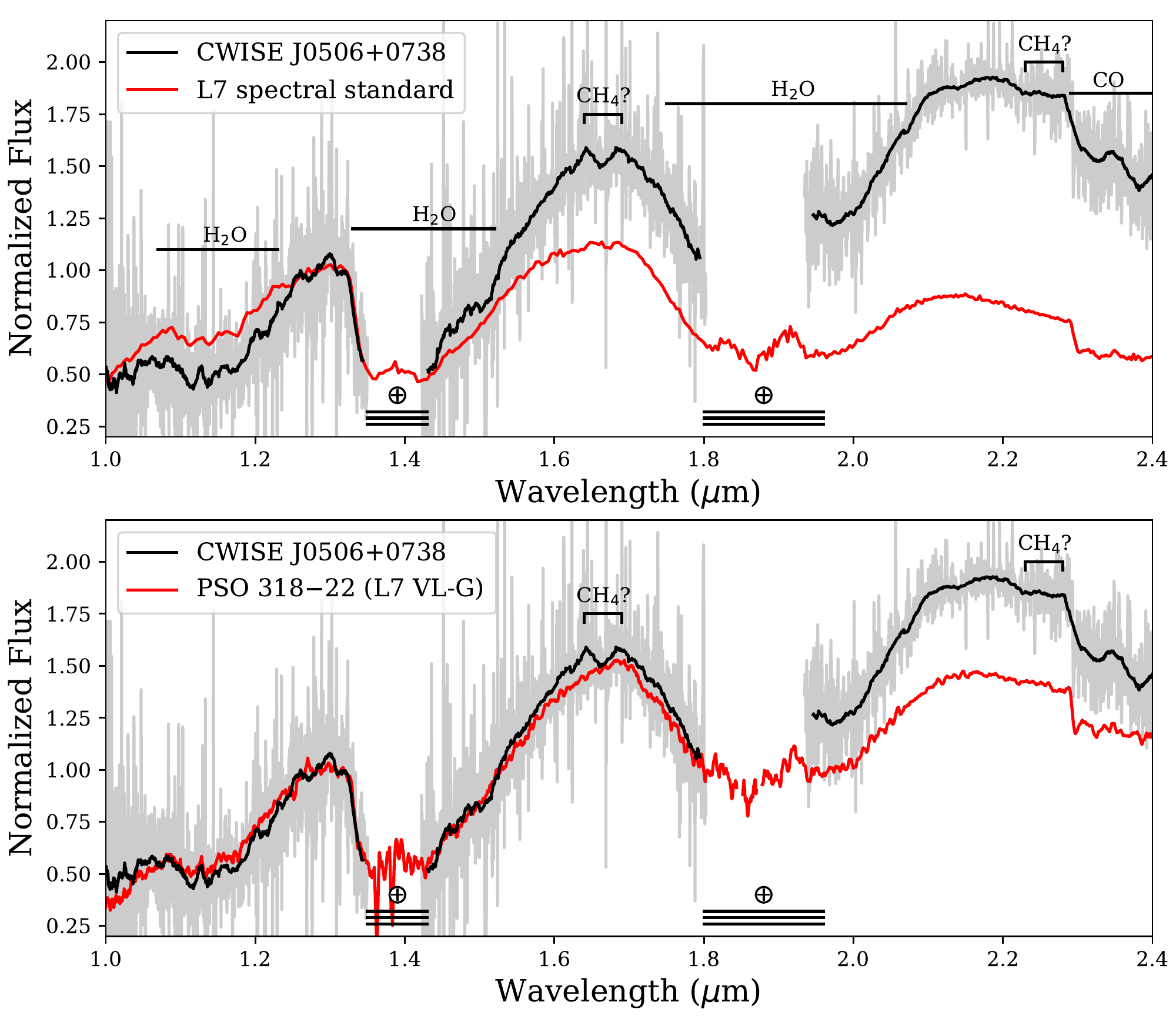}
\caption{The Keck/NIRES spectrum of CWISE J0506$+$0738, shown
in the original resolution (grey lines) and smoothed to a resolution of $\lambda/\Delta\lambda$ $\approx$ 100 (black lines).
CWISE J0506$+$0738 is compared to the L7 spectral standard 2MASSI J0825196$+$211552 \citep{kirkpatrick2000, cruz2018} in the top panel, and the young L7 VL-G dwarf PSO J318.5338$-$22.8603 \citep{liu2013} in the bottom panel. Both comparisons highlight the extremely red nature of CWISE J0506$+$0738. All spectra are normalized between 1.27 and 1.29 $\mu$m, and prominent absorption features have been labeled.  
}  
\label{fig:spectrum}
\end{figure*}

The inter-band flux calibration for Keck/NIRES orders is occasionally skewed by seeing or differential refraction slit losses.  In particular, there is a gap between the third ($K$-band) and fourth ($H$-band) orders spanning 1.86 to 1.89 $\mu$m\footnote{https://www2.keck.hawaii.edu/inst/nires/genspecs.html}, and the overlap between the fourth and fifth ($J$-band) orders lies in a region of strong telluric and stellar H$_2$O absorption.  We therefore re-scaled the resulting spectrum to have a $J-K$ synthetic color consistent with UKIRT $J$-band and UHS $K$-band photometry by applying small multiplicative constants to the $H$- and $K$-band portions of the spectrum.  The final reduced spectrum is shown in Figure \ref{fig:spectrum}.

\section{Analysis}
\label{sec:anal}

\subsection{Spectral Type}
\label{sec:spt}
As with many of the known, young, late-type red L dwarfs, none of the L dwarf spectral standards \citep{kirkpatrick2010, cruz2018} provide a suitable match to the near-infrared spectrum of CWISE J0506$+$0738.  The best match to the $J$-band portion of the spectrum is the L7 standard 2MASSI J0825196$+$211552 \citep{kirkpatrick1999,cruz2018}, which is shown in the top panel of Figure \ref{fig:spectrum}.  CWISE J0506$+$0738 shows much stronger H$_2$O absorption around 1.1 $\mu$m, a feature commonly seen in low-gravity L dwarfs.  This comparison also shows how red CWISE J0506$+$0738 is compared to a normal, field-age/field-gravity late-L dwarf.  The bottom panel of Figure \ref{fig:spectrum} shows a comparison of CWISE J0506$+$0738 with PSO J318.5338$-$22.8603 \citep{liu2013}, which is typed as L7 VL-G in that work.  These two objects match relatively well across the $J$-band portion of the spectrum, though the extreme redness of CWISE J0506$+$0738 can still be seen in this comparison via the mismatch in the $H$- and $K$-band portions of their spectra.  

\begin{figure*}
\plotone{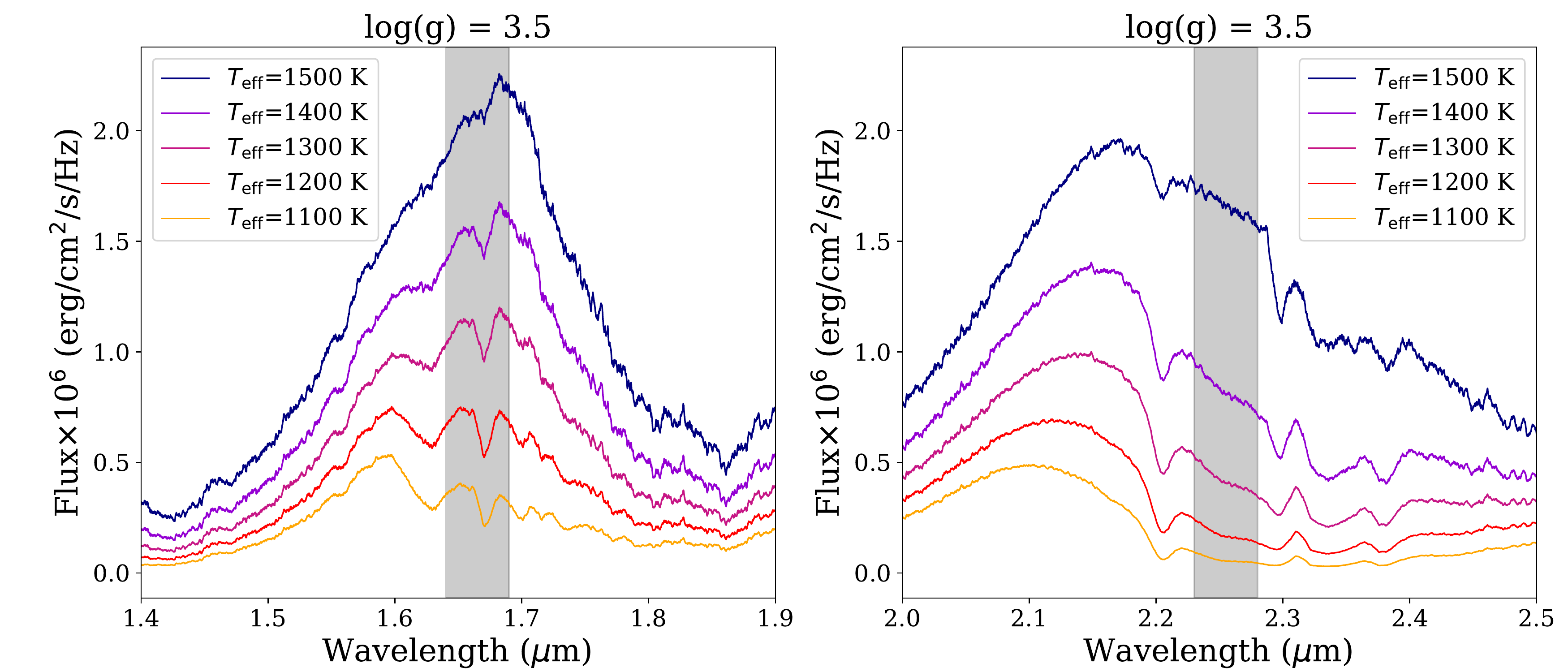}
\caption{Model spectra from \cite{marley2021} with varying effective temperatures and surface gravity fixed at log(g)=3.5.  The gray bands highlight the approximate regions of the absorption features seen in the spectrum of CWISE J0506$+$0738.  }  
\label{fig:ch4}
\end{figure*}

We also note that the spectrum of CWISE J0506$+$0738 has a noticeable absorption feature at the $H$-band peak. There is also a second, less-pronounced absorption feature present in the $K$-band portion of CWISE J0506$+$0738's spectrum between 2.2 and 2.3 $\mu$m.  While we cannot {\em a priori} rule out systematic noise or a data reduction artifact for these features, we note that no similar features have been seen in Keck/NIRES spectra of L dwarfs obtained and reduced by our group (e.g., \citealt{meisner2021, schapera2022, softich2022, theissen2022}).  We also note that these features occur at the approximate locations of CH$_4$ absorption seen in model spectra of low-surface gravity brown dwarfs with effective temperatures $\lesssim$1400 K. Figure \ref{fig:ch4} compares solar-metallicity model spectra from \cite{marley2021} with fixed low-surface gravities (log(g)=3.5) and varying effective temperatures.  Prominent methane absorption features can be seen in the $H$- and $K$-bands for \teff\ $\lesssim$1400 K.  While these models are informative for (potentially) identifying the source of some of the absorption features seen in the spectrum of CWISE J0506$+$0738, we were unable to find any models that successfully reproduced the overall shape of CWISE J0506$+$0738's spectrum, similar to previous studies of young brown dwarfs (e.g., \citealt{manjavacas2014}).

The presence of CH$_4$ in the $H$- and $K$-band peaks of CWISE J0506$+$0738's spectrum would suggest that this source is early T dwarf \citep{burgasser2006}, although these features are fairly weak in strength.  \cite{charnay2018} showed that the presence of clouds can greatly reduce the abundance of CH$_4$ in the photospheres of low-gravity objects, a possible explanation for the absence of CH$_4$ bands in the spectra of 2M1207b and HR8799bcd \citep{barman2011a, barman2011b, konopacky2013}. If the same effect holds here, it would argue for a particularly low temperature for CWISE J0506$+$0738, below that of the {\teff} $\approx$ 1200~K planetary-mass L dwarf PSO J318.5338$-$22.8603 and VHS 1256$-$1257B which originally showed no indication of CH$_4$ absorption in the 1--2.5~$\mu$m region\footnote{Recent high S/N {\em JWST}/NIRSPEC observations of VHS~1256$-$1257B have revealed the presence of weak 1.6~$\micron$ absorption in its spectrum \citep{miles2022}.}. \citep{liu2013, gauza2015}. These two sources do have detectable absorption in the 3.3 $\mu$m $\nu_3$ CH$_4$ fundamental band \citep{miles2018}, and cloud scattering opacity is likely responsible for muting the 1.6~$\mu$m and 2.2~$\mu$m bands in these red L dwarfs \citep{charnay2018,burningham2021}.  Indeed, it has been noted previously that PSO J318.5338$-$22.8603 is just on the warmer side of the transition to CH$_4$ becoming the dominant carbon-bearing molecule in its atmosphere \citep{tremblin2017}.  We tentatively assert that both $H$- and $K$-band features in the spectrum of CWISE J0506$+$0738 are due to CH$_4$ absorption, which may be tested with more detailed analysis (e.g., atmospheric retrievals; \citealt{burningham2017, burningham2021}) and higher S/N moderate-resolution data.  Given the similarity of the $J$-band portion of CWISE J0506$+$0738's spectrum to PSO J318.5338$-$22.8603 (L7 VL-G), and likely detection of CH$_4$ in the $H$- and $K$-bands, we assign a near-infrared spectral type of L8$\gamma$--T0$\gamma$ to CWISE J0506$+$0738, where the $\gamma$ signifies very low surface gravity \citep{kirkpatrick2005}.

\subsection{Spectral Evidence of Youth}
\label{sec:youth}
The characterization of brown dwarfs and planetary mass objects as ``low surface gravity'' or ``young'' typically arises from gravity-sensitive (or more specifically, photosphere pressure-sensitive) spectral features quantified by spectral indices (e.g., \citealt{steele1995, martin1996, luhman1997, gorlova2003, mcgovern2004, kirkpatrick2006, allers2007, manjavacas2020}).  Many of these spectral indices, however, are designed for optical spectra (e.g., \citealt{cruz2009}) or are only applicable to objects with spectral types earlier than $\sim$L5 (e.g., \citealt{allers2013, lodieu2018}).  The $H$-cont index is a gravity-sensitive index defined in \cite{allers2013} that is one of the few gravity-sensitive indices applicable to spectral types later than L5.  This index is designed to approximate the slope of the blue side of the $H$-band peak, with low-gravity objects exhibiting a much steeper slope than field-age brown dwarfs.  However, this index is defined using a band centered at 1.67 $\mu$m, which is where a feature potentially attributable to CH$_4$ occurs in our spectrum.  Thus the $H$-cont index does not provide an accurate assessment of the slope of the blue side of the $H$-band peak for this object.  

\begin{figure}
\plotone{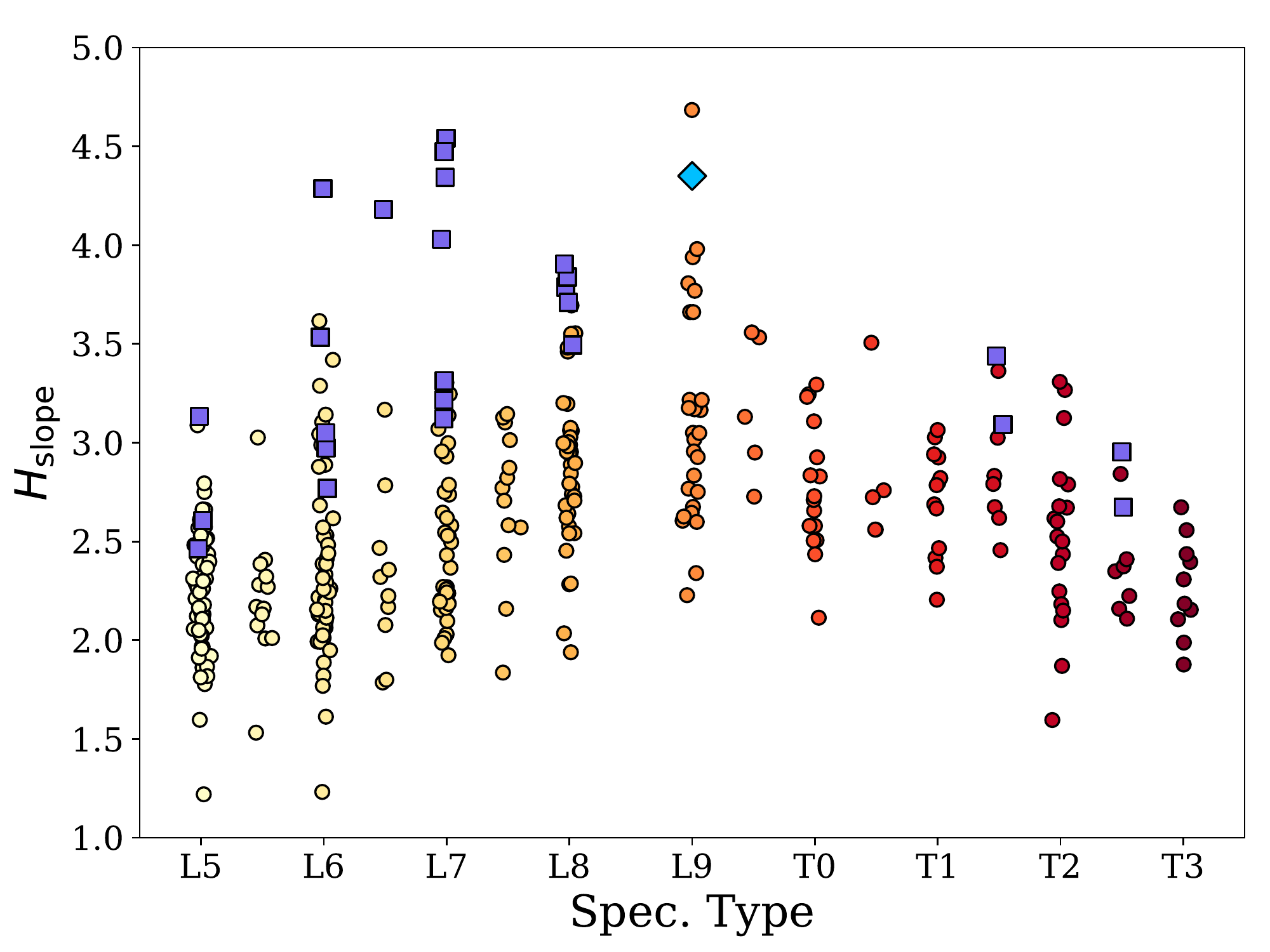}
\caption{$H$-band slope index versus spectral type for field late-L and T dwarfs (colored circles) based on data from the SPLAT archive \citep{burgasser2017}, with colors corresponding to spectral type.  Young L and T dwarfs are represented by purple squares.  CWISE J0506$+$0738 (blue diamond) is an outlier amongst field-age late-Ls, similar to the young, late-type L dwarf population. Small offsets have been added to spectral type values for differentiation purposes.}  
\label{fig:slope-index}
\end{figure}

We have created a modified slope index for the blue side of the $H$-band peak by computing a simple linear least-squares fit to the 1.45--1.64 $\mu$m region after normalizing to the $J$-band peak between 1.27 and 1.29 $\mu$m.  We measured this slope (normalized flux/$\mu$m) for several late-L and early-T dwarfs, both field and young association members, as shown in Figure~\ref{fig:slope-index}.  We note that the largest slope for the entire sample belongs to WISE J173859.27$+$614242.1, an object that has been difficult to classify \citep{mace2013}, but is most consistent with an extremely red L9 \citep{thompson2013}.  It is unclear if this object is young, has an extremely dusty photosphere, or both.  For typical L7-T0 dwarfs, $H$-slope values for field objects range from 2--4, while equivalently classified young L dwarfs have values that range over 3--5. For CWISE J0506$+$0738, we find a slope of 4.38, significantly larger than field-age late-L dwarfs.   The known population of young, very red L dwarfs similarly has larger $H$-slope values than their field-age counterparts.  

\cite{schneider2014} also showed that the H$_2$($K$) index defined in \cite{canty2013} could distinguish young, low-gravity late-Ls from the field late-L population. The H$_2$($K$) index determines the slope of the $K$-band between 2.17~$\mu$m and 2.24~$\mu$m.  CWISE J0506$+$0738 has an H$_2$($K$) value of 1.030, which is again consistent with the known population of low-gravity late-type L dwarfs (1.029 $\leq$ H$_2$($K$) $\leq$ 1.045) compared field-age L6--L8 brown dwarfs (H$_2$($K$) $\gtrsim$ 1.05).

Another spectral feature that has been used to distinguish low-surface gravity late-L dwarfs are the K I absorption lines between 1.1 and 1.3 $\mu$m \citep{mcgovern2004,allers2013,miles2022}.  Our Keck/NIRES spectrum does not have sufficient S/N around the $J$-band peak to investigate these lines.  A higher S/N spectrum would help to ensure no ambiguity regarding the surface gravity of CWISE J0506$+$0738.

\subsection{Radial Velocity}
\label{sec:rv}

The resolution of the Keck/NIRES data is sufficient to obtain a coarse measure of the radial velocity (RV) of CWISE J0506$+$0738, particularly in the vicinity of strong molecular features. We followed a procedure similar to that described in \citep{burgasser2015} (see also \citealt{blake2010,hsu2021}), forward-modeling the wavelength-calibrated spectrum prior to telluric correction in the 2.26--2.38~$\mu$m region.  This spectral band contains the prominent 2.3~$\mu$m CO 2-0 band present in L dwarf spectra, as well as strong telluric features that allow refinement of the spectral wavelength calibration (cf. \citealt{newton2014}). We used a {\teff} = 1300~K, {\logg} = 4.5~dex (cgs) BTSettl atmosphere model ($M[\lambda]$) from \citet{allard2012} which provides the best match to the CO band strength, and a telluric absorption model ($T[\lambda]$) from \citet{livingston1991}. We forward modeled the data ($D[\lambda]$) using four parameters: the barycentric radial velocity of the star (RV$_\oplus$), the strength of telluric absorption ($\alpha$), the instrumental gaussian broadening profile width ($\sigma_{broad}$), and the wavelength offset from the nominal SpeXtool solution ($\Delta\lambda$): 
\begin{equation}
    D[\lambda] = \left(M[\lambda^*+\Delta\lambda]\times{T[\lambda+\Delta\lambda]^\alpha}\right) \\ \otimes\kappa_G(\sigma_{broad})
\end{equation}
with $\lambda^* = \lambda(1+{RV_\oplus}/{c})$ accounting for the radial motion of the star and $\kappa_G$ representing the gaussian broadening kernel.  Preliminary fits that additionally included rotational broadening of the stellar spectrum indicated that this parameter was equal to the instrumental broadening and is likely unresolved ($v\sin{i}$ $\lesssim$ 65~km/s), so it was ignored in our final fit.

\begin{figure*}
\plotone{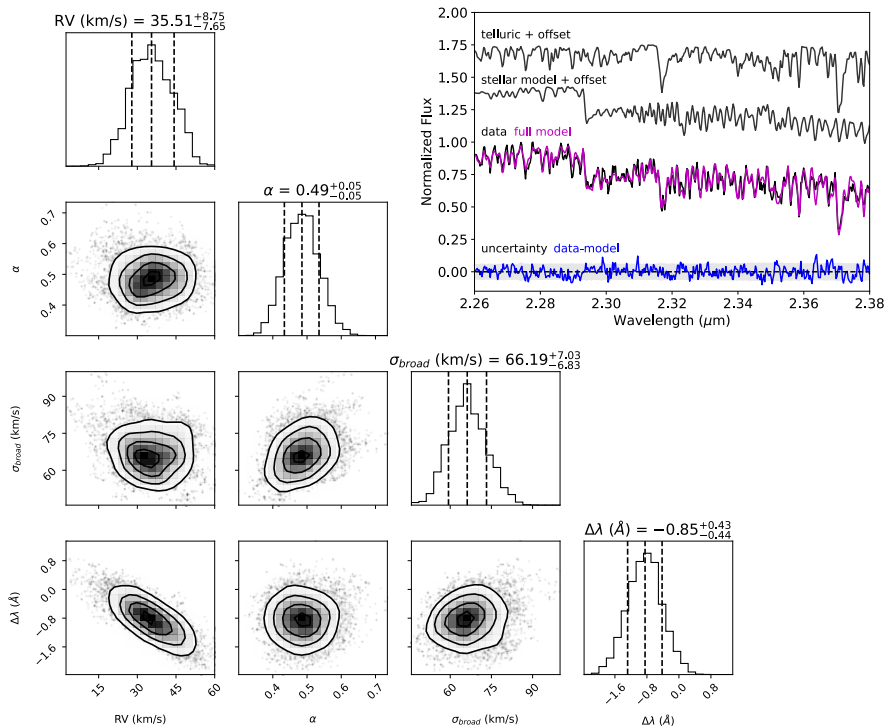}
\caption{MCMC forward model fit of the normalized 2.26--2.38~$\mu$m spectrum of CWISE J0506$+$0738 for RV measurement. The panels along the diagonal show the posterior distributions for our four fitting parameters: the barycentric radial velocity of the star (RV$_\oplus$ in km/s), the strength of the telluric absorption ($\alpha$), the instrumental gaussian broadening profile width ($\sigma_{broad}$ in km/s), and the wavelength offset from the nominal SpeXtool solution ($\Delta\lambda$ in {\AA}). The lower left panels illustrate correlations between parameters; only the RV and $\Delta\lambda$ parameters show a modest inverse correlation, effectively expanding the uncertainty on the RV measurement.  The upper right corner shows the NIRES spectrum of CWISE J0506$+$0738 prior to telluric correction (black line) and the best-fit model spectrum (magenta line) composed of stellar model and telluric absorption components (offset lines above fit). Residuals (data minus model, blue line) are consistent with measurement uncertainties (grey band).  
}  
\label{fig:rv}
\end{figure*}

After an initial ``by-eye'' optimization of parameters, we used a simple Markov Chain Monte Carlo (MCMC) algorithm to explore the parameter space, evaluating goodness of fit between model and data using a $\chi^2$ statistic. Figure~\ref{fig:rv} displays the posterior distribution of our fit parameters after removing the first half of the MCMC chain (``burn-in''), which are normally distributed. There is a small correlation between RV$_\oplus$ and $\Delta\lambda$ which is expected given that stellar and telluric features are intermixed in this region. This correlation increases the uncertainties of these parameters. We find that the best-fit model from this analysis is an excellent match to the NIRES spectrum, with residuals consistent with uncertainties. After correction for barycentric motion ($-$19.2~km/s), we determine a heliocentric radial velocity of +16.3$^{+8.8}_{-7.7}$~km s$^{-1}$ for for CWISE J0506$+$0738.  

\begin{figure*}
\plotone{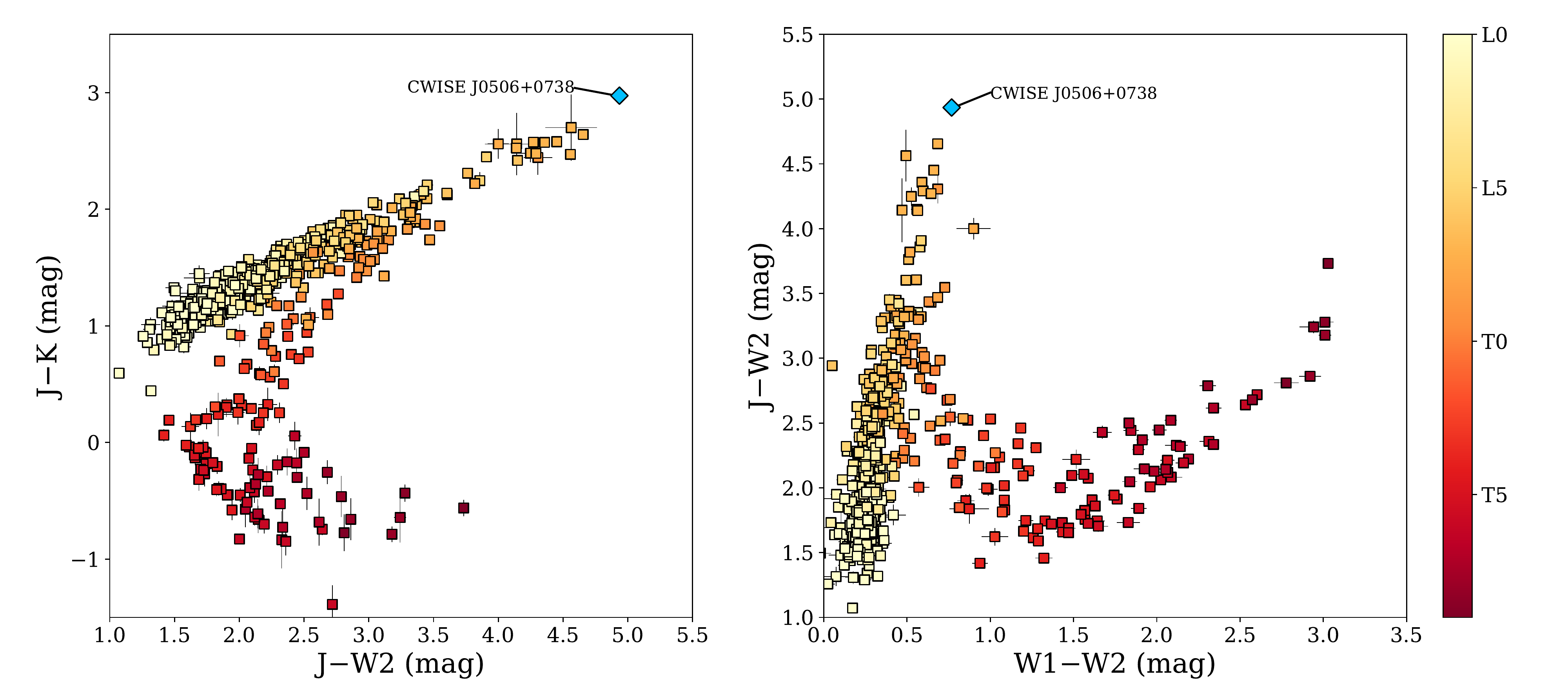}
\caption{Color-color diagrams showing known brown dwarfs recovered in the UKIRT Hemisphere Survey (Schneider et al.~in prep), supplemented with known red L dwarfs from Table \ref{tab:redLs}.  CWISE J0506$+$0738 is a clear outlier, being significantly redder than other known L dwarfs both in $J-K$ and $J-$W2 color. }  
\label{fig:ccds}
\end{figure*}

\section{Discussion}
\label{sec:discussion}

\subsection{Redder than Red}
\label{sec:red}

CWISE J0506$+$0738 has exceptionally red colors compared to the known brown dwarf population.  Figure \ref{fig:ccds} highlights this by comparing CWISE J0506$+$0738 to other UHS DR2 L and T dwarfs (Schneider et al.~in prep.) and red L dwarfs not covered by the UHS survey.   Table \ref{tab:redLs} summarizes photometric and spectral type information for all known free-floating L dwarfs with $J-K$ colors greater than 2.2 mag.  All photometry is on the MKO system and comes from the VISTA Hemisphere Survey (VHS; \citealt{mcmahon2013}), \cite{liu2016}, or \cite{best2021}.  WISE J173859.27$+$614242.1 has no near-infrared MKO photometry in the literature or in available catalogs.  For this source, we used its low-resolution near-infrared spectrum published in \citet{mace2013} normalized to its most precise $K$-band photometric measurement (2MASS $K_{\rm S}$; \citealt{skrutskie2006}), and then computed synthetic $J_{\rm MKO}$ and $K_{\rm MKO}$ photometry.  Even amongst known red L dwarfs, CWISE J0506$+$0738 stands out as exceptionally red, being $\sim$0.3 mag redder in both $(J-K)_{MKO}$ and $J_{MKO}-$W2 color than all other known free-floating L dwarfs.    

Directly imaged planetary-mass companions also have exceptionally red near-infrared colors. Some of the L-type companions (Table~\ref{tab:redLs}) do not have {\em WISE} W1 (3.4 $\mu$m) and W2 (4.6 $\mu$m) photometry, but have equivalent Spitzer/IRAC photometry in ch1 (3.6 $\mu$m) and ch2 (4.5 $\mu$m). For HD 203030B, we use $J$- and $K$-band photometry from \citet{metchev2006} and \citet{miles2017}, and convert Spitzer/IRAC ch1 and ch2 photometry from \cite{martinez2022} using the Spitzer-WISE relations from \cite{kirkpatrick2021}.  For VHS 1256$-$1257B, we use $J$- and $K$-band photometry from \cite{gauza2015}, and convert Spitzer/IRAC ch2 photometry from \cite{zhou2020} to W2 using the \cite{kirkpatrick2021} relation.  We chose not to use the published W1 photometry of VHS 1256$-$1257B from \cite{gauza2015} because of its large uncertainty (0.5 mag).  For BD$+$60 1417B, all photometry comes directly from \cite{faherty2021}.  Both HD 203030B and BD$+$60 1417B are included in both panels of Figure \ref{fig:ccds}, while VHS 1256$-$1257B is included in the left panel of Figure \ref{fig:ccds}. We note that none of these companions have $(J-K)_{MKO}$ or $J_{MKO}-$W2 colors as red as CWISE J0506$+$0738.  Of the remaining planetary-mass companions that lack 3--5~$\mu$m photometry, only 2M1207b ($J-K$=3.07$\pm$0.23 mag; \citealt{chauvin2004, chauvin2005, mohanty2007, patience2010}) and HD 206893B ($J-K$=3.36$\pm$0.08 mag; \citealt{milli2017, delorme2017, krammerer2021, meshkat2021, ward2021}) have redder $J-K$ colors than CWISE J0506$+$0738.

\subsection{WISE Photometric Variability}

Young brown dwarfs have been shown to have enhanced photometric variability compared to field-age brown dwarfs \citep{biller2015, metchev2015, schneider2018, vos2020, vos2022}.  Most brown dwarfs with detected variability at 3--5 $\mu$m, measured largely with Spitzer/IRAC, have amplitudes of a few percent or less (see compilation in \citealt{vos2020}). Multi-epoch photometry from WISE generally does not have the precision to detect such variability (\citealt{mace2015}, Brooks et al.~submitted).  However, objects with extremely high-amplitude variability could be distinguished in multi-epoch WISE data.    

\begin{figure*}
\plotone{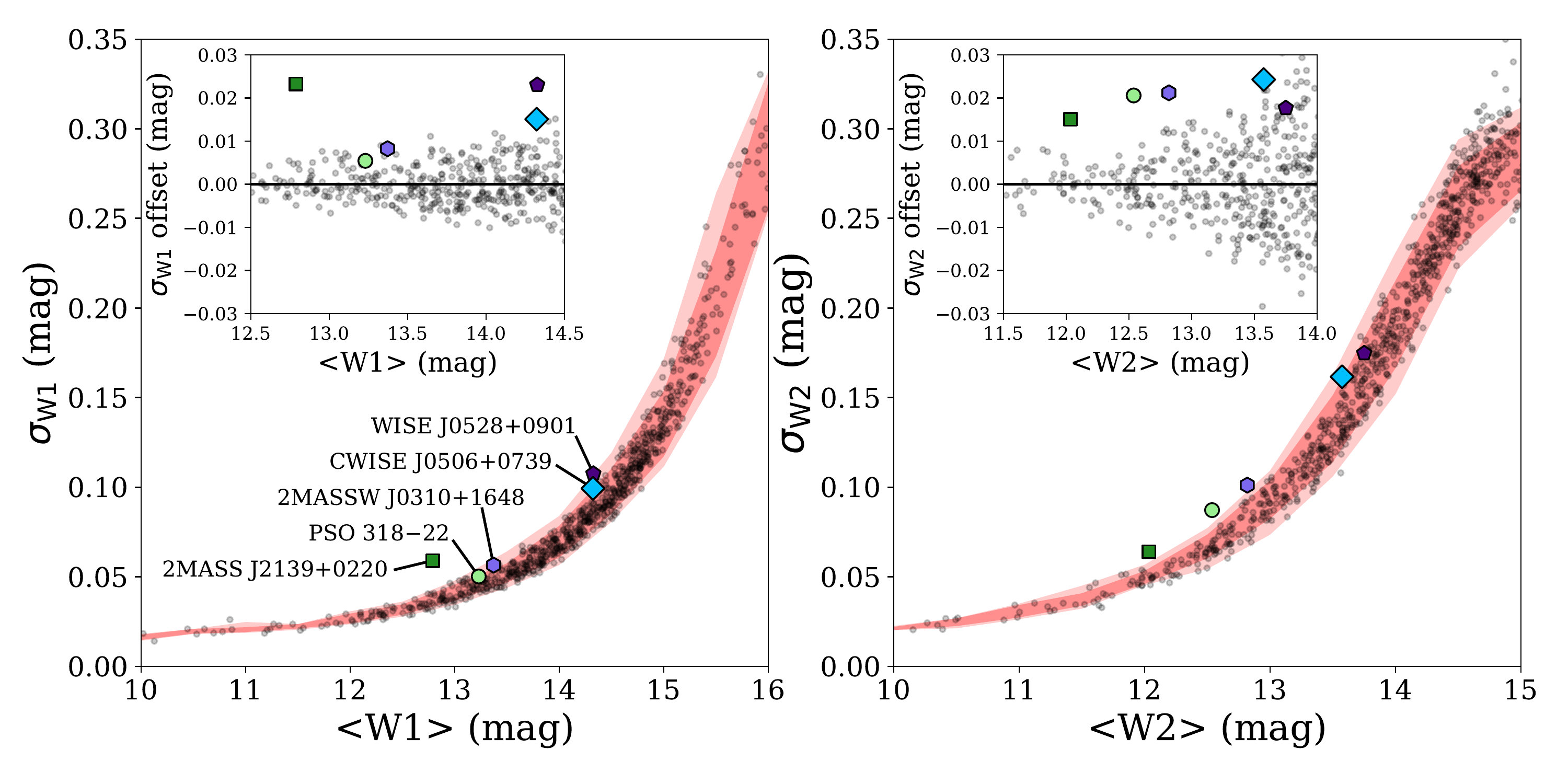}
\caption{Standard deviation ($\sigma$) versus average magnitude over all single-exposure WISE/NEOWISE W1 (left) and W2 (right) detections of known brown dwarfs. Color contours indicate 16--84\% and 5--95\% confidence intervals in 0.5 magnitude bins.  The insets on each panel show the difference between measured $\sigma$ values and polynomial fits to the magnitude trend. 2MASS J2139$+$0220 (dark green square), PSO J318.5338$-$22.8603 (light green circle), 2MASSW J0310599$+$164816 (light purple hexagon), CWISE J0506$+$0738 (cyan diamond), and WISE J052857.68$+$090104.4 (dark purple pentagon) are all highlighted as clear deviants from these trends. }  
\label{fig:var}
\end{figure*}

Given tentative evidence of near-infrared photometric variability (see Section \ref{sec:ukirt}), we investigated WISE \citep{wright2010} and NEOWISE \citep{mainzer2011, mainzer2014} data for evidence of mid-infrared variability for CWISE J0506$+$0738.  WISE/NEOWISE has been scanning the mid-infrared sky for over 10 years, and a typical location on the sky has been observed with the W1 and W2 filters every six months since early 2010.\footnote{With the exception of a $\sim$3 year gap between the initial WISE mission and reactivation as NEOWISE from February 2011 to December 2013.}  During each $\sim$1 day visit, 10--15 individual exposures are typically acquired.  We chose to analyze these single exposures as opposed to epochal coadds (e.g. ``unTimely''; \citealt{meisner2022}) because CWISE J0506$+$0738 is brighter than the nominal threshold where single exposure photometry becomes unreliable, especially at W2 ($\sim$14.5 mag; \citealt{schneider2016a}); and the concern that the coadded frames would dilute any traces of photometric variability.  Such coadded photometry may prove useful for future investigations of long-term/long-period variability. 

We gathered photometry from the WISE/NEOWISE Single Exposure Source Catalogs \citep{wise2020a, wise2020b, wise2020c, neowise2020} for CWISE J0506$+$0738 and the same set of known L, T, and Y dwarfs shown in Figure \ref{fig:ccds}.  Collectively, these objects should have comparable levels of low-amplitude variability generally undetectable by WISE.  For each source, we measured the average and standard deviation of both W1 and W2 magnitudes.  We omit frames with {\it qual\_frame} values equal to zero, as these frames likely have contaminated flux measurements.  Because single exposure frames are subject to astronomical transients (e.g., cosmic ray hits, satellite streaks), we excluded 4$\sigma$ outliers from the set of single exposure photometry for each source.  We also excluded sources that were either blended or contaminated (e.g., bright star halos, diffraction spikes). 

Figure \ref{fig:var} compares mean and standard deviation values, which show clear trends in both W1 and W2 photometry.  We immediately identify four objects with magnitudes between 12 and 14.5 that have photometric scatter above the 5--95\% confidence interval ($\gtrsim$2$\sigma$) in either W1 or W2. 

\noindent
{\em 2MASS 21392676$+$0220226 (2MASS J2139$+$0220)} is a T1.5 dwarf \citep{burgasser2006} that is well-known  for its large-amplitude infrared variability.  \cite{radigan2012} monitored 2MASS J2139$+$0220 and found $J$-band variability with a peak-to-peak amplitude of $\sim$26\%, which until recent observations of VHS 1256$-$1257B \citep{zhou2022} was the highest amplitude variability found for any brown dwarf.  Since the \cite{radigan2012} study, this object has been the subject of numerous variability investigations \citep{apai2013, khandrika2013, karalidi2015}, with \cite{yang2016} finding variability of 11--12\% in Spitzer/IRAC ch1 and ch2 photometry.  The extreme variability of 2MASS J2139$+$0220 is attributed to variations in the thickness of silicate clouds \citep{apai2013, karalidi2015, vos2022b}.  This object has also been shown to have a nearly edge-on inclination \citep{vos2017}, and is a kinematic member of the $\sim$200 Myr-old Carina-Near moving group \citep{zhang2021}.  

\noindent
{\em  WISE J052857.68$+$090104.4 (WISE~J0528$+$0901)} is a clear W1 outlier, originally classified as a late-M giant by \cite{thompson2013} but later reclassified as a very low-gravity L1  brown dwarf member of the $\sim$20 Myr 32 Orionis group \citep{burgasser2016}.  This planetary-mass object has an anomalous $J-$W2 color, suggestive of excess flux at 5 $\mu$m, although \cite{burgasser2016} found no evidence of circumstellar material or cool companions.  The source may also be a variable in the W2 band, but its fainter magnitude here makes it less distinct than comparably bright L and T dwarfs. Nevertheless, these data suggest that WISE~J0528$+$0901 has an unusually dusty and variable atmosphere, making it a compelling source for future photometric monitoring. 

\noindent
{\em PSO J318.5338$-$22.8603} is a clear W2 outlier and exceptionally red $\beta$ Pic member that has been shown to have large-amplitude infrared variability in the infrared \citep{biller2015, vos2019}, with a peak-to-peak amplitude of 3.4\% in Spitzer/IRAC ch2 photometry \citep{biller2018}.  Interestingly, PSO J318.5338$-$22.8603 is an outlier in W2 and not in W1, which may indicate a cloud depth effects given that the W1 and W2 bands probe different depths in the atmosphere.  

\noindent
{\em 2MASSW J0310599$+$164816 (2MASS~J0310+1648AB)} is another W2 outlier, and is an optically classified L8 \cite{kirkpatrick2000}.  This object is a resolved (0\farcs2) $\sim$equal brightness binary \citep{stumpf2010} that shows evidence of high amplitude variability in the near-infrared \citep{buenzli2014}.  While the variability observations were not long enough to determine a true amplitude or period, the brightening rate of $\sim$2\% per hour was the largest measured in the sample. While there is no clear evidence of youth for 2MASS~J0310+1648AB in the literature, this object was typed as L9.5 (sl.~red) in \cite{schneider2014}. Further investigation of the potential youth and cloud properties of this object may be warranted.

CWISE J0506$+$0738 joins this group of variability outliers, as one of very few objects with both W1 and W2 scatter outside the 16--84\% confidence interval of comparable-brightness L and T dwarfs.  To estimate the amplitude of variability associated with these deviations, we fit tenth-order polynomials to the scatter versus magnitude trends in W1 and W2, and calculated RMS values by finding the magnitude offset (in quadrature) for our outlying targets. Assuming sinusoidal variability, RMS values can be converted to peak-to-peak amplitudes with a multiplicative factor of 2$\sqrt{2}$.  Using the 16--84\% confidence region as uncertainties for the predicted values from the polynomial fits, we find peak-to-peak variability on the order of 13$\pm$1\% for W1 and 12$\pm$2\% for W2 for 2MASS J2139$+$0220, which is generally consistent with results from Spitzer \citep{yang2016}. For CWISE J0506$+$0738, we estimate  15$\pm$5\% variability for W1 and 23$\pm$9\% variability for W2.  Variability at these levels would certainly be extraordinary; however, we caution that the relatively low precision of WISE/NEOWISE single exposure measurements may inflate these results.  Future photometric and/or spectroscopic monitoring would help to explore the variability properties of CWISE J0506$+$0738. 

\subsection{Distance}
\label{sec:dist}

CWISE J0506$+$0738 is faint at optical wavelengths and was therefore undetected by the Gaia mission \citep{gaia2022}.  The currently available astrometry for CWISE J0506$+$0738 is insufficient for a parallax measurement. Because CWISE J0506$+$0738 has such an unusually shaped spectrum, standard spectral-type versus absolute magnitude relations for normal, field-age brown dwarfs are not applicable.  There have been efforts to create relations between absolute magnitudes and spectral types for low-gravity brown dwarfs; however, these are typically valid for spectral types earlier than L7 (e.g., \citealt{faherty2016, liu2016}).  \cite{faherty2013} found absolute photometry of the young L5 dwarf 2MASS J03552337$+$1133437 was fainter than field L5 dwarfs at wavelengths shorter than $\sim$2.5 $\mu$m, and brighter at longer wavelengths. \cite{schneider2016b} investigated other young, red L dwarfs with measured parallaxes and found that $K$-band photometry produced photometric distances that aligned well with parallactic distances.  This trend was also noted in \cite{filippazzo2015}, \cite{faherty2016}, and \cite{liu2016}.

\begin{figure}
\plotone{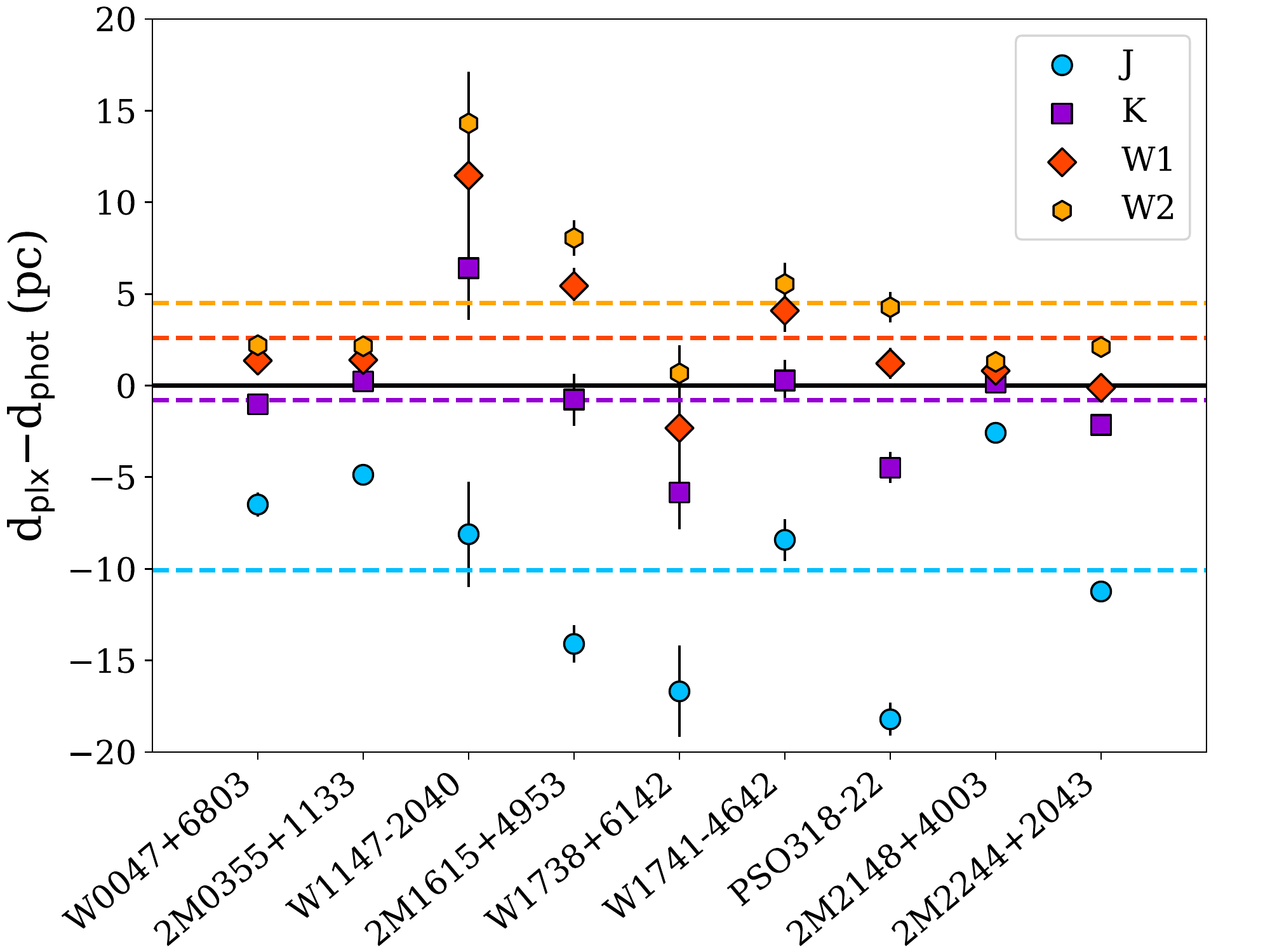}
\caption{A comparison of photometric and parallactic distances for free-floating objects from Table \ref{tab:redLs} with measured parallaxes.  Objects are labeled on the x-axis.  Dashed lines show average differences between photometric and parallactic distances for each band, with colors corresponding to those given in the legend. }  
\label{fig:dist}
\end{figure}

Here, we use nine young, free-floating brown dwarfs (Table \ref{tab:redLs}) with measured parallaxes \citep{liu2016, best2020, kirkpatrick2021, gaia2022} to compare measured distances to photometric distances based on absolute magnitude-spectral type relations for $J_{\rm MKO}$, $K_{\rm MKO}$, W1, and W2 (\citealt{dupuy2012,kirkpatrick2021}; Figure~\ref{fig:dist}). Consistent with prior results, we find that $K_{\rm MKO}$-band photometric distances (average offset $\Delta$d = $-$0.8~pc, scatter $\sigma_d$ = 3.3~pc) are generally more accurate than $J_{\rm MKO}$ ($\Delta$d = $-$10~pc, $\sigma_d$ = 5.1~pc), W1 ($\Delta$d = +2.6~pc, $\sigma_d$ = 3.8~pc), or W2 ($\Delta$d = +4.5~pc, $\sigma_d$ = 4.1~pc) photometric distances.  To ensure these values are not biased, we also evaluated the fractional difference for each photometric band, defined as $\Delta$d/d$_{\rm plx}$, and find that $K$-band photometric distances are typically within 5\% for this sample, compared to 52\%, 11\%, and 20\% for $J_{\rm MKO}$,  W1, and W2, respectively.

Using the absolute magnitude-spectral type relation from \cite{dupuy2012}, a spectral type of L9$\pm$1, and its measured $K_{MKO}$ photometry, we estimate a photometric distance of 32$^{+4}_{-3}$ pc for CWISE J0506$+$0738.  Again, given the exceptional nature of this source, and its unknown multiplicity, we advise that this distance estimate be used with caution until it can be confirmed with a trigonometric parallax.   

\subsection{Moving Group Membership}
\label{sec:mg}

Young brown dwarfs are often associated both spatially and kinematically with young, nearby moving groups, thereby serving as invaluable age benchmarks.

To assess the potential moving group membership of CWISE J0506$+$0738, we use the BANYAN $\Sigma$ algorithm \citep{gagne2018}, which deploys a Bayesian classifier to assign probabilities of moving group membership through 6D coordinate alignment (position and velocity) to 26 known moving groups in the solar neighborhood.  We used the position and proper motion of CWISE J0506$+$0738 from UKIRT and UHS measurements (Table \ref{tab:cwise0506}), and our measured radial velocity from the NIRES spectrum (Section \ref{sec:rv}). With these values alone, we find an 82\% membership probability in the $\beta$ Pictoris moving group (BPMG; \citealt{zuckerman2001}), a 3\% membership probability in the AB Doradus moving group (ABDMG; \citealt{zuckerman2004}), and a 15\% probability of being unassociated with any moving group.  The predicted/optimal distances for membership in BPMG and ABDMG are 32~pc and 64~pc, respectively; our estimated distance clearly aligns with the former.  If we include the distance estimate in the BANYAN $\Sigma$ algorithm, the probability of BPMG membership goes up to 99\%. 

We also tested the kinematic membership of CWISE J0506$+$0738 using the LACEwING analysis code \citep{riedel2017}.  Again, using just the position, proper motion, and radial velocity of CWISE J0506$+$0738, we find non-zero probabilities for ABDMG (56\%), the Argus Moving Group (71\%), BPMG (28\%), the Columba Association (52\%), and the Tucana-Horologium Association (6\%).  Note that LACEwING is stricter in assigning membership probabilities than BANYAN, with bona fide BPMG members having a maximum membership probability of $\sim$70\% when only proper motion and radial velocity are used \citep{riedel2017}.  If we use our photometric distance as an additional constraint, BPMG is returned as the group with the highest probability of membership at 86\%.  

Membership in the $\beta$ Pictoris moving group is clearly favored for CWISE J0506$+$0738, although a directly measured distance is necessary for confirmation.  If confirmed, CWISE J0506$+$0738 would have the latest spectral type and lowest mass amongst  free-floating BPMG members, following PSO J318.5338$-$22.8603 \citep{liu2013}.  Several candidate members with L7 or later spectral types have also been proposed (\citealt{best2015, schneider2017, kirkpatrick2021, zhang2021}; however, see \citealt{hsu2021}).  PSO J318.5338$-$22.8603 has proven to be an exceptionally valuable laboratory for studying planetary-mass object atmospheres \citep{biller2015, biller2018, allers2016, faherty2016}.  A second planetary-mass object in this group that bridges the L/T transition will further contribute to these studies.

Assuming $\beta$ Pic membership, we can use the group age of 22$\pm$6 Myr \citep{shkolnik2017} to estimate the mass of CWISE J0506$+$0738.  To do this, we must first estimate the luminosity ($L_{\rm bol}$) or effective temperature (\teff) of the source.  For the former, we used the empirical $K$-band bolometric correction/spectral type relation for young brown dwarfs quantified in \cite{filippazzo2015}. Combining this with the UHS $K$-band magnitude and our distance estimate, we infer a bolometric luminosity of $\log$($L_{\rm bol}$/$L_{\odot}$) = -4.55$\pm$0.12.  We caution that this value is based on our estimated distance from Section \ref{sec:dist}, and will need to be updated when a measured parallax becomes available. We then used the solar metallicity evolutionary models of \cite{marley2021} to infer a mass of 7$\pm$2 $M_{\rm Jup}$.  The evolutionary models also provide a radius of 1.32$\pm$0.03 $R_{\rm Jup}$ for these parameters, consistent with the radii of low-gravity late-type L dwarfs \citep{filippazzo2015}.  Combining this radius with our bolometric luminosity, we find \teff\ = 1140$\pm$80 K.  This is $\sim$130 K cooler than a field-age L9 \citep{kirkpatrick2021}, consistent with previous works showing low-gravity late-Ls tend to be $\sim$100--200 K cooler than field-age objects at the same spectral type \citep{filippazzo2015, faherty2016}.  In particular, this temperature is 50-100~K cooler than \teff\ estimates of PSO J318.5338$-$22.8603 \citep{liu2013,miles2018}, consistent with the appearance of CH$_4$ absorption at lower temperatures.  

The predicted mass of 7$\pm$2 $M_{\rm Jup}$ is well below the deuterium-fusion minimum mass of 14 $M_{\rm Jup}$ commonly used to distinguish brown dwarfs from planetary mass objects. As such, this object helps bridge the mass gap between the lowest mass free-floating $\beta$ Pic members and directly imaged exoplanets, such as 51 Eri b ($\sim$T6.5; \citealt{macintosh2015, rajan2017}). CWISE J0506$+$0738 could also help to constrain the effective temperature of the L/T transition at an age of $\sim$20-25 Myr \citep{binks2014, bell2015, messina2016, nielsen2016, shkolnik2017, miret2020}.  CWISE J0506$+$0738 would be one of the youngest objects to join a small but growing number of benchmark substellar objects with known ages a the L/T transition such as HD 203030B (30--150 Myr; \citealt{metchev2006, miles2017}), 2MASS J13243553+6358281 ($\sim$150 Myr; \citealt{looper2007, gagne2018b}), HIP 21152B and other T-type Hyades members ($\sim$650 Myr; \citealt{kuzuhara2022, schneider2022}), $\epsilon$ Indi Ba ($\sim$3.5 Gyr; \citealt{scholz2003, chen2022}), and the white dwarf companion COCONUTS-1 ($\sim$7 Gyr; \citealt{zhang2020}).

\section{Summary}

We have presented the discovery and analysis of an exceptionally red brown dwarf, CWISE J0506$+$0738, identified as part of the Backyard Worlds: Planet 9 citizen science project.  The near-infrared spectrum of CWISE J0506$+$0738 is highly reddened and shows signatures of low-surface gravity, as well as weak absorption features that we associate with methane bands. This object has the reddest $J-K$ and $J-$W2 colors of any free-floating L-type brown dwarf, and we tentatively assign a near-infrared spectral type of L8$\gamma$--T0$\gamma$. The exceptionally red color of CWISE J0506$+$0738 may be due to several factors. Objects with low surface gravities have inefficient gravitational settling of silicate dust grains, which can remain high in the atmospheres.  Such grains can be directly detected at long wavelengths (e.g., \citealt{cushing2006, burgasser2008, suarez2022}) and could be constrained for CWISE J0506$+$0738 with future long-wavelength observations (e.g., \citealt{miles2022}).  The angle at which a brown dwarf is viewed has also been shown to affect its near-infrared colors, with objects viewed equator-on tending to have redder colors than those viewed pole-on \citep{vos2017}. A measurement of CWISE J0506$+$0738's rotational period combined with its rotational velocity (e.g., $v$sin$i$) from a high-resolution spectrum could determine whether or not CWISE J0506$+$0738 is viewed closer to pole-on or equator-on.   A high-resolution spectrum would also allow for a higher precision radial velocity measurement and a more detailed probe of gravity-sensitive features.

CWISE J0506$+$0738's astrometry and kinematics points to likely membership in the 22~Myr $\beta$ Pictoris moving group, to be confirmed or rejected with future trigonometric parallax and higher precision radial velocity measurements.  If associated, CWISE J0506$+$0738 would be the lowest-mass $\beta$ Pictoris member found to date, with an estimated mass of 7$\pm$2 $M_{\rm Jup}$, well within the planetary-mass regime. The extreme colors of this object, and its relatively low proper motion ($<$100 mas yr$^{-1}$), suggests the existence of a other extremely red L dwarfs that may have been missed by previous searches due to assumptions about brown dwarf colors or selection requirements for large proper motions. Recent large-scale near-infrared surveys such as UHS \citep{dye2018} and VHS \citep{mcmahon2013} that push several magnitudes deeper than previous efforts (e.g., 2MASS) may be able to confidently detect the faint $J$-band magnitudes of similar objects. 

Because of this object's unique spectroscopic properties, and the fact that young brown dwarfs often display large-amplitude variability (e.g., \citealt{vos2022}), CWISE J0506$+$0738 is an intriguing target for future photometric or spectroscopic variability monitoring.  Longer wavelength observations with the James Webb Space Telescope would have the additional advantage of further constraining the existence and abundance of CH$_4$ and analyzing the presence and properties of dust grains through silicate absorption features \citep{miles2022}.

\begin{longrotatetable} 
\begin{deluxetable*}{lcccccccccccc}
\label{tab:redLs}
\tablecaption{Infrared Photometry for L Dwarfs with $J-K$ $>$ 2.2 mag}
\tablehead{
\colhead{Name} & \colhead{Disc.} & \colhead{SpT} & \colhead{SpT} & \colhead{$J_{\rm MKO}$} & \colhead{$K_{\rm MKO}$} & \colhead{NIR} & \colhead{W1} & \colhead{W2} & \colhead{$(J-K)_{\rm MKO}$} & \colhead{$J_{\rm MKO}-$W2}\\
\colhead{} & \colhead{Ref.} & \colhead{} & \colhead{Ref.} & \colhead{(mag)} & \colhead{(mag)} & \colhead{Ref.} & \colhead{(mag)} & \colhead{mag} & \colhead{(mag)} & \colhead{(mag)}  }
\startdata
\cutinhead{Free Floating}
WISEP J004701.06$+$680352.1 & 1 & L6--L8$\gamma$ & 2 & 15.490$\pm$0.070 & 13.010$\pm$0.030 & 3 & 11.768$\pm$0.010 & 11.242$\pm$0.008 & 2.480$\pm$0.076 & 4.248$\pm$0.070 \\
PSO 057.2893$+$15.2433 & 4 & L7 red & 4 & 17.393$\pm$0.027 & 14.869$\pm$0.012 & 20 & 13.818$\pm$0.014 & 13.254$\pm$0.012 & 2.524$\pm$0.030 & 4.139$\pm$0.030  \\
2MASS J03552337$+$1133437 & 5 & L3--L6$\gamma$ & 2 & 13.940$\pm$0.003 & 11.491$\pm$0.001 & 20 & 10.617$\pm$0.012 & 10.032$\pm$0.008 & 2.449$\pm$0.003 & 3.908$\pm$0.009 \\
CWISE J050626.96$+$073842.4 & 6 & L8--T0$\gamma$ & 6 & 18.487$\pm$0.017 & 15.513$\pm$0.022 & 6,20 & 14.320$\pm$0.015 & 13.552$\pm$0.013 & 2.974$\pm$0.028 & 4.935$\pm$0.021 \\
WISEA J090258.99$+$670833.1 & 7 & L7 red & 7 & 16.864$\pm$0.246 & 14.305$\pm$0.108 & 8 & 13.192$\pm$0.013 & 12.722$\pm$0.009 & 2.559$\pm$0.269 & 4.142$\pm$0.246  \\
2MASS J11193254$-$1137466 & 9 & L7 VL-G\tablenotemark{a} & 10 & 17.330$\pm$0.029 & 14.751$\pm$0.012 & 21 & 13.540$\pm$0.014 & 12.879$\pm$0.010 & 2.580$\pm$0.032 & 4.451$\pm$0.031 \\
WISEA J114724.10$-$204021.3 & 11 & L7$\gamma$ & 12 & 17.445$\pm$0.028 & 14.872$\pm$0.011 & 21 & 13.677$\pm$0.013 & 13.088$\pm$0.011 & 2.573$\pm$0.030 & 4.357$\pm$0.030 \\
2MASS J16154255$+$4953211 & 13 & L3--L6$\gamma$ & 2 &  16.506$\pm$0.016 & 14.260$\pm$0.070 & 3,20 & 13.225$\pm$0.012 & 12.648$\pm$0.008 & 2.246$\pm$0.072 & 3.858$\pm$0.018\\
WISE J173859.27$+$614242.1 & 14 & L9 pec(red) & 14 & 17.680$\pm$0.110\tablenotemark{c} & 15.237$\pm$0.100\tablenotemark{c} & 6,22 & 14.059$\pm$0.011 & 13.374$\pm$0.009 & 2.443$\pm$0.149\tablenotemark{c} & 4.306$\pm$0.100\tablenotemark{c}\\
WISE J174102.78$-$464225.5 & 15 & L5--L7$\gamma$ & 2 & 15.951$\pm$0.010 & 13.533$\pm$0.005 & 21 & 12.362$\pm$0.027 & 11.802$\pm$0.024 & 2.418$\pm$0.011 & 4.149$\pm$0.026\\
PSO J318.5338$-$22.8603 & 16 & L7 VL-G & 16 & 17.181$\pm$0.018 & 14.540$\pm$0.009 & 21 & 13.210$\pm$ 0.013 & 12.526$\pm$0.010 & 2.640$\pm$0.020 & 4.655$\pm$0.021\\
2MASS J21481628$+$4003593 & 17 & L6.5 pec & 17 & 14.054$\pm$0.003 & 11.745$\pm$0.001 & 20 & 10.801$\pm$0.011 & 10.292$\pm$0.007 & 2.309$\pm$0.003 & 3.762$\pm$0.008 \\
ULAS J222711$-$004547 & 18 & L7 pec & 18 & 17.954$\pm$0.039 & 15.475$\pm$0.014 & 21\tablenotemark{b} & 14.259$\pm$0.014 & 13.663$\pm$0.013 & 2.479$\pm$0.041 & 4.291$\pm$0.041\\
2MASS J22443167$+$2043433 & 19 & L6--L8$\gamma$ & 2 & 16.401$\pm$0.016 & 13.826$\pm$0.006 & 20 & 12.775$\pm$0.012 & 12.130$\pm$0.008 & 2.575$\pm$0.017 & 4.271$\pm$0.018\\
\cutinhead{Companions}
BD$+$60 1417B & 23 & L6--L8$\gamma$ & 23 & 18.53$\pm$0.20 & 15.83$\pm$0.20 & 23 & 14.461$\pm$0.014 & 13.967$\pm$0.013 & 2.70$\pm$0.28 & 4.46$\pm$0.20 \\
HD 203030B & 24 & L7.5 & 24 & 18.77$\pm$0.08 & 16.21$\pm$0.10 & 24,25 & 15.67$\pm$0.02\tablenotemark{d} & 14.77$\pm$0.02\tablenotemark{d} & 2.56$\pm$0.13 & 4.00$\pm$0.08 \\
VHS 1256$-$1257B & 26 & L7.5 & 26 & 17.136$\pm$0.020 & 14.665$\pm$0.010 & 21 & \dots & 12.579$\pm$0.020\tablenotemark{e} & 2.471$\pm$0.022 & 4.557$\pm$0.028 \\
2MASS J1207334$-$393254b & 27,28 & L3 VL-G & 29 & 20.0$\pm$0.2 & 16.93$\pm$0.11 & 27,30 & \dots & \dots & 3.07$\pm$0.23 & \dots \\
HD 206893B & 31 & L4--L8 & 32 & 18.38$\pm$0.03 & 15.02$\pm$0.07 & 32 & \dots & \dots & 3.36$\pm$0.08 & \dots \\
2MASS J22362452+4751425b & 33 & late-L pec & 33 & 19.97$\pm$0.11 & 17.28$\pm$0.04 & 33 & \dots & \dots & 2.69$\pm$0.12 & \dots \\
HR 8799b & 34 & L5--T2 & 35 & 19.46$\pm$0.17 & 16.99$\pm$0.06 & 36,37,38 & \dots & \dots & 2.47$\pm$0.18 & \dots \\
\enddata
\tablenotetext{a}{2MASS J11193254$-$1137466 is a binary \citep{best2017} and the spectral type listed is the unresolved spectral type.}
\tablenotetext{b}{ULAS J222711$-$004547 also has $J$- and $K$-band photometry in the UKIRT Large Area Survey (LAS; \citealt{lawrence2007}).  We use the VHS photometric measurements here because they have smaller uncertainties than those in the UKIRT LAS.}
\tablenotetext{c}{Near-infrared photometry for WISE J173859.27$+$614242.1 was determined synthetically from its near-infrared spectrum.}
\tablenotetext{d}{Converted from Spitzer ch1 and ch2 photometry in \cite{miles2017} using relations in \cite{kirkpatrick2021}.}
\tablenotetext{e}{Converted from Spitzer ch2 photometry in \cite{zhou2020} using relations in \cite{kirkpatrick2021}.}
\tablerefs{(1) \cite{gizis2012}; (2) \cite{gagne2015}; (3) \cite{liu2016}; (4) \cite{best2015}; (5) \cite{reid2006}; (6) This work; (7) \cite{schneider2017}; (8) \cite{best2021}; (9) \cite{kellogg2015}; (10) \cite{best2017}; (11) \cite{schneider2016b}; (12) \cite{faherty2016}; (13) \cite{metchev2008}; (14) \cite{mace2013}; (15) \cite{schneider2014}; (16) \cite{liu2013b}; (17) \cite{looper2008}; (18) \cite{marocco2014}; (19) \cite{dahn2002}; (20) UHS (\citealt{dye2018}, Bruursema et al.~in prep.); (21) VHS \citep{mcmahon2013}; (22) 2MASS \citep{skrutskie2006}; (23) \cite{faherty2021}; (24) \cite{metchev2006}; (25) \cite{miles2017}; (26) \cite{gauza2015};   (27) \cite{chauvin2004}; (28) \cite{chauvin2005}; (29) \cite{allers2013}; (30) \cite{mohanty2007}; (31) \cite{milli2017}; (32) \cite{ward2021}; (33) \cite{bowler2017}; (34) \cite{marois2008}; (35) \cite{bowler2010}; (36) \cite{esposito2013}; (37) \cite{oppenheimer2013}; (38) \cite{liu2016} }
\end{deluxetable*}
\end{longrotatetable}

\acknowledgments

.

The Backyard Worlds: Planet 9 team would like to thank the many Zooniverse volunteers who have participated in this project. We would also like to thank the Zooniverse web development team for their work creating and maintaining the Zooniverse platform and the Project Builder tools. This research was supported by NASA grant 2017-ADAP17-0067.  This material is supported by the National Science Foundation under Grant No. 2007068, 2009136, and 2009177.  This publication makes use of data products from the UKIRT Hemisphere Survey, which is a joint project of the United States Naval Observatory, The University of Hawaii Institute for Astronomy, the Cambridge University Cambridge Astronomy Survey Unit, and the University of Edinburgh Wide-Field Astronomy Unit (WFAU).  UHS is primarily funded by the United States Navy.  The WFAU gratefully acknowledges support for this work from the Science and Technology Facilities Council through ST/T002956/1 and previous grants.  The authors acknowledge the support provided by the US Naval Observatory in the areas of celestial and reference frame research, including the USNO's postdoctoral program.  (Some of) The data presented herein were obtained at the W. M. Keck Observatory, which is operated as a scientific partnership among the California Institute of Technology, the University of California and the National Aeronautics and Space Administration. The Observatory was made possible by the generous financial support of the W. M. Keck Foundation.  This publication makes use of data products from the {\it Wide-field Infrared Survey Explorer}, which is a joint project of the University of California, Los Angeles, and the Jet Propulsion Laboratory/California Institute of Technology, and NEOWISE which is a project of the Jet Propulsion Laboratory/California Institute of Technology. {\it WISE} and NEOWISE are funded by the National Aeronautics and Space Administration.  Part of this research was carried out at the Jet Propulsion Laboratory, California Institute of Technology, under a contract with the National Aeronautics and Space Administration.  This publication makes use of data products from the Two Micron All Sky Survey, which is a joint project of the University of Massachusetts and the Infrared Processing and Analysis Center/California Institute of Technology, funded by the National Aeronautics and Space Administration and the National Science Foundation.  This research has benefitted from the SpeX Prism Spectral Libraries, maintained by Adam Burgasser.  The authors wish to recognize and acknowledge the very significant cultural role and reverence that the summit of Maunakea has always had within the indigenous Hawaiian community.  We are most fortunate to have the opportunity to conduct observations from this mountain.

\facilities{UKIRT/WFCAM, Keck/NIRES, WISE, NEOWISE}

\software{
BANYAN~$\Sigma$ \citep{gagne2018}, 
CASUTOOLS \citep{irwin2004}, 
LACEwING \citep{riedel2017}, 
SpeXTool \citep{cushing2004}, 
SPLAT \citep{burgasser2014},
WiseView \citep{caselden2018}
}

\end{document}